\documentclass[twocolumn,aps,pra,longbibliography,10pt]{revtex4-1}

\usepackage{amsmath,amssymb,graphicx,color,bm}
\usepackage[utf8]{inputenc}
\usepackage{ulem}

\newcommand{\real}{\mathrm{Re}}

\newcommand{\sech}{\text{sech}}

\newcommand{\commentout}[1]{}

\begin{document}

\title{Overcoming dispersive spreading of quantum wave packets\\ via periodic nonlinear kicking}

\author{Arseni Goussev$^1$, Phillipp Reck$^{2,3}$, Florian Moser$^2$, Antonio Moro$^1$, Cosimo Gorini$^2$, Klaus Richter$^2$}

\affiliation{$^1$Department of Mathematics, Physics and Electrical Engineering, Northumbria University, Newcastle Upon Tyne NE1 8ST, United Kingdom\\
$^2$Institut f\"ur Theoretische Physik, Universit\"at Regensburg, D-93040 Regensburg, Germany\\
$^3$SPEC, CEA, CNRS, Universit\'e Paris-Saclay, CEA-Saclay, 91191 Gif-sur-Yvette Cedex, France}

\begin{abstract}
We propose the suppression of dispersive spreading of wave packets governed by the free-space Schr\"odinger equation with a periodically pulsed nonlinear term. 
Using asymptotic analysis, we construct stroboscopically-dispersionless quantum states that are physically reminiscent of, but mathematically different from, 
the well-known one-soliton solutions of the nonlinear Schr\"odinger equation with a constant (time-independent) nonlinearity. 
Our analytics are strongly supported by full numerical simulations. 
The predicted dispersionless wave packets can move with arbitrary velocity and can be realized in experiments involving ultracold atomic gases with temporally controlled interactions. 
\end{abstract}

\date{\today}

\maketitle


\section{Introduction}

As time elapses, the wave function representing a freely-propagating nonrelativistic quantum particle inevitably changes its shape -- a process commonly referred to as dispersive spreading. The spreading of a free-space wave function can be entirely suppressed only if one gives up the normalization condition. Thus, in one spatial dimension, there exist only two types of nonnormalizable quantum waves -- the plain wave and the Airy packet -- that preserve their shape during the free motion \cite{BB79Nonspreading, Gre80Comment, UR96Uniqueness}. Such waves however do not represent the probability density of a single quantum particle, but should rather be regarded as describing a statistical ensemble of infinitely many free particles \cite{BB79Nonspreading}.

Physicists' desire to tame dispersive spreading of normalizable wave packets is as old as quantum mechanics itself. Schr\"odinger was the first to find an example of a quantum system -- a particle trapped inside a stationary harmonic potential -- that supports localized wave packets moving without dispersion along classical trajectories \cite{Sch26stetige}. To date, the existence of nonspreading wave packets has been established in a wide range of physical systems and lies at the heart of a vibrant area of research on (generalized) coherent states, much of it reviewed in Refs.~\cite{ZFG90Coherent, DM03Theory, Gaz09Coherent, DodCoherent}.
Here we briefly discuss two specific classes which are relevant for the central result of this work. 
The first class encompasses periodically driven quantum systems, in which dispersive wave packet spreading is suppressed via creation of a nonlinear resonance between 
an internal oscillatory mode of the system and an external periodic driving. 
The underlying quantum evolution remains linear at all times, and the nondispersive wave packets appear as localized eigenstates 
of the corresponding Floquet Hamiltonian (see Ref.~\cite{BDZ02Non-dispersive} for a review).  
The second class includes nonlinear quantum systems, i.e. the ones governed by the nonlinear Schr\"odinger equation (NLSE), 
that support nondispersive wave packets commonly referred to as solitons \cite{KM89Dynamics,APTNLS}. 
In these systems, the nonlinear term in the evolution equation is essential to overcome dispersive spreading 
and to allow the wave packet to propagate without deformation.

\begin{figure}[htb!]
	\centering
	\includegraphics[width=0.4\textwidth]{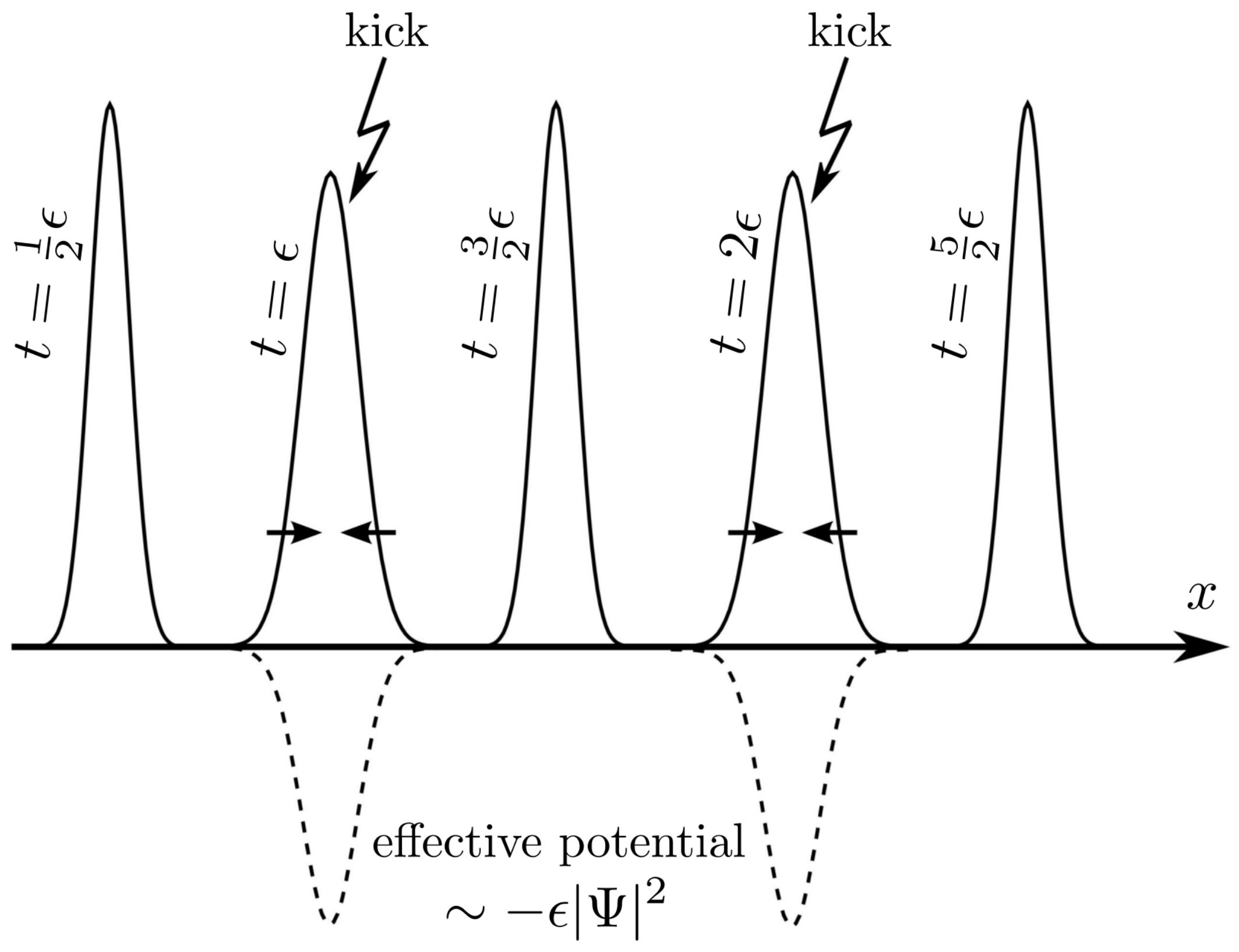}
	\caption{Sketch of a stroboscopic soliton, the solution to Eq.~(\ref{NLSE-dimless}). Solid curves represent the probability distribution $|\Psi(x,t)|^2$ at the kick instants (integer $t/\epsilon$) and half way between the adjacent kicks (half-integer $t/\epsilon$). Dashed curves illustrate the effective potential caused by the corresponding kick. Short arrows point in the direction of the classical ``squeezing force'' due to the effective potential.} 
	\label{fig:sketch}
\end{figure}

The two classes are largely nonoverlapping.  
Rare exceptions are systems subjected to Feshbach resonance management \cite{kevrekidis2003}, 
a technique proposed for atom optics experiments requiring a periodical sign change of the interparticle interaction strength
\footnote{According to the botanics of Ref.~\cite{bukov2015}, such systems belong to the ``Kapitza class''}.
In this paper we address a related class
of quantum systems that, loosely speaking, can be placed in between that of linear Floquet and nonlinear time-independent Schr\"odinger systems. 
More specifically, the quantum dynamics introduced here consists of long intervals of linear free-particle motion 
interspersed periodically with short (near-instantaneous) intervals of nonlinear evolution (see Fig.~\ref{fig:sketch} for an illustration).
We demonstrate that such periodic nonlinear kicking is sufficient for certain families of localized normalizable wave packets 
to overcome dispersive spreading. When observed stroboscopically at the kicking frequency these wave packets retain their shape 
and propagate with a constant (but arbitrary) velocity, thus behaving as ``stroboscopic solitons.''
The latter arise as solutions of a nonlinear integral equation -- Eq.~\eqref{eq-phi-lin_op} below -- which, 
to the best of our knowledge, has not been considered before.
While our analytics predicts the existence of stroboscopic soliton solutions in an asymptotic regime of weak kicking,
our numerics show their remarkable robustness even away from it.
From the physical point of view, 
the robustness implies that their experimental realization is well within state-of-the-art cold atom techniques
\cite{Clark2015,nguyen2017,everitt2017, arunkumar2018}.
On the purely mathematical side, proving the existence (and form) of exact stroboscopic solutions represents
a new open problem.


\section{Problem formulation}

We consider the motion of a one-dimensional quantum particle of mass $m$ described by the wave function $\tilde{\Psi}(\tilde{x}, \tilde{t})$, with $\tilde{x}$ and $\tilde{t}$ denoting the space and time variables, respectively. The time evolution of the wave function is governed by the NLSE, in which the nonlinear (Kerr-type) term is switched on and off periodically:
\begin{equation}
	i \frac{\partial \tilde{\Psi}}{\partial \tilde{t}} = -\frac{\hbar}{2 m} \frac{\partial^2 \tilde{\Psi}}{\partial \tilde{x}^2} - \lambda \!\!\! \sum_{n = -\infty}^{\infty} \delta(\tilde{t} - n T) |\tilde{\Psi}|^2 \tilde{\Psi} \,.
\label{NLSE-dim}
\end{equation}
Here $\delta(\cdot)$ denotes the Dirac delta function, and $\lambda > 0$ and $T$ represent the strength and period of the nonlinear kicking, respectively. 
The wave function is normalized to unity, $\int_{-\infty}^{+\infty} d\tilde{x} \, |\tilde{\Psi}|^2 = 1 \,$.
It is convenient to rewrite Eq.~\eqref{NLSE-dim} in a dimensionless form by introducing
\begin{equation}
	x = \frac{m \lambda}{\hbar T} \tilde{x} \,, \quad t = \frac{m \lambda^2}{\hbar T^2} \tilde{t} \,, \quad \Psi(x,t) = \sqrt{\frac{\hbar T}{m \lambda}} \tilde{\Psi}(\tilde{x}, \tilde{t}) \,.
\label{rescaling}
\end{equation}
In terms of the new variables, the NLSE becomes
\begin{equation}
	i \frac{\partial \Psi}{\partial t} = -\frac{1}{2} \frac{\partial^2 \Psi}{\partial x^2} - \epsilon \!\!\! \sum_{n = -\infty}^{+\infty} \delta(t - n \epsilon) |\Psi|^2 \Psi
\label{NLSE-dimless}
\end{equation}
with a single dimensionless parameter
\begin{equation}
\label{God_parameter}
	\epsilon = \frac{m \lambda^2}{\hbar T} \,,
\end{equation}
and normalization $\int_{-\infty}^{+\infty} dx \, |\Psi|^2 = 1$.
It is worth emphasizing the simplicity of our setup: the protocol involves piecewise linear, free evolution,
i.e.~no further specific background potentials are required.

We now look for stroboscopic soliton solutions to Eq.~(\ref{NLSE-dimless}) under the constrain imposed by normalization. That is, we aim to find a wave function $\psi(x) = \Psi(x, 0^+)$, describing the system immediately after the kick at time $t = 0$, such that $\Psi(x,\epsilon^+)$, corresponding to the instant immediately after the kick at $t = \epsilon$, coincides with $\psi(x)$, modulo an overall spatial displacement and a global (position-independent) phase shift. In other words, we want to find all complex functions $\psi(x)$, velocities $v$ and angular frequencies $\omega$ that satisfy the equation
\begin{equation}
	U^{\epsilon} \psi(x) = e^{i \omega \epsilon} \psi(x - v \epsilon) \,,
\label{main_eq}
\end{equation}
where $U^{\epsilon}$ denotes the time-evolution operator propagating $\Psi(x, 0^+)$ into $\Psi(x,\epsilon^+)$. It follows from Eq.~(\ref{NLSE-dimless}) (see Appendix~\ref{app:A}) that $U^{\epsilon} = K^{\epsilon} U_0^{\epsilon}$,
where $U_0^{\epsilon}$ is the evolution operator for the corresponding linear Schr\"odinger equation, $i \frac{\partial \Psi}{\partial t} = -\frac{1}{2} \frac{\partial^2 \Psi}{\partial x^2}$, and the operator $K^{\epsilon}$ describes an instantaneous transformation of the wave function caused by a nonlinear kick. Hence, for an arbitrary normalizable wave function $f(x)$, one has
\begin{align}
	U_0^{\epsilon} f(x)
	&= \exp \left( \frac{i \epsilon}{2} \frac{\partial^2}{\partial x^2} \right) f(x) \\
	&= \sqrt{\frac{1}{2 \pi i \epsilon}} \int_{-\infty}^{+\infty} dy \, \exp \left( i \frac{(x - y)^2}{2 \epsilon} \right) f(y)
\end{align}
and $K^{\epsilon} f(x) = f(x) \exp\left( i \epsilon |f(x)|^2 \right)$.

Introducing a new complex function $\phi(x)$ according to
\begin{equation}
	\psi(x) = e^{i v x} \phi(x) \,,
\label{gauge}
\end{equation}
we rewrite Eq.~(\ref{main_eq}) as (see Appendix~\ref{app:B})
\begin{equation}
	U^{\epsilon} \phi(x) = e^{i \alpha \epsilon} \phi(x) \,, \qquad \alpha = \omega - \frac{v^2}{2} \,.
\label{eq-phi}
\end{equation}
Equation~(\ref{gauge}) plays the role of a gauge transformation, and $\phi(x)$ describes a stationary stroboscopic soliton of frequency $\alpha$ and
normalization $\int_{-\infty}^{+\infty} dx \, |\phi|^2 = 1$.

\section{Asymptotic analysis}

We proceed with constructing an asymptotic solution to the problem defined by Eq.~(\ref{eq-phi}) and $\int_{-\infty}^{+\infty} dx \, |\phi|^2 = 1$ in the limit of small $\epsilon$. It proves convenient to first transform Eq.~(\ref{eq-phi}) into a nonlinear integral equation with a linear integral operator. Written explicitly, Eq.~(\ref{eq-phi}) reads
\begin{equation}
	e^{i \epsilon |U_0^\epsilon \phi|^2} U_0^{\epsilon} \phi = e^{i \alpha \epsilon} \phi \,.
\end{equation}
It follows immediately that $|U_0^{\epsilon} \phi| = |\phi|$, and thus
\begin{equation}
	U_0^{\epsilon} \phi = \phi e^{i \epsilon \left( \alpha - |\phi|^2 \right)} \,.
\label{eq-phi-lin_op}
\end{equation}
We now look for a solution to Eq.~(\ref{eq-phi-lin_op}) in the form of a power series in $\epsilon$:
\begin{equation}
	\phi = \phi_0 + \epsilon \phi_1 + \epsilon^2 \phi_2 + O(\epsilon^3) \,.
	\label{phi_expansion_def}
\end{equation}
Substituting the series into Eq.~(\ref{eq-phi-lin_op}), and expanding it in powers of $\epsilon$, 
we obtain to leading order (see Appendix~\ref{app:C})
\begin{equation}
	-\tfrac{1}{2} \phi_0'' + \left( \alpha - |\phi_0|^2 \right) \phi_0 = 0 \,.
\label{phi_0_eq}
\end{equation}
Similarly, expanding the normalization integral in powers of $\epsilon$, we get to leading order $\int_{-\infty}^{+\infty} dx \, |\phi_0|^2 = 1$.
Equation~(\ref{phi_0_eq}) admits the leading order soliton solution
\begin{equation}
	\phi_0 = \frac{1}{2} \, \sech \frac{x}{2}
	\label{phi_0-sol}
\end{equation}
with $\alpha=1/8$.

\begin{figure*}[ht!]
	\centering
	\includegraphics[width=\textwidth]{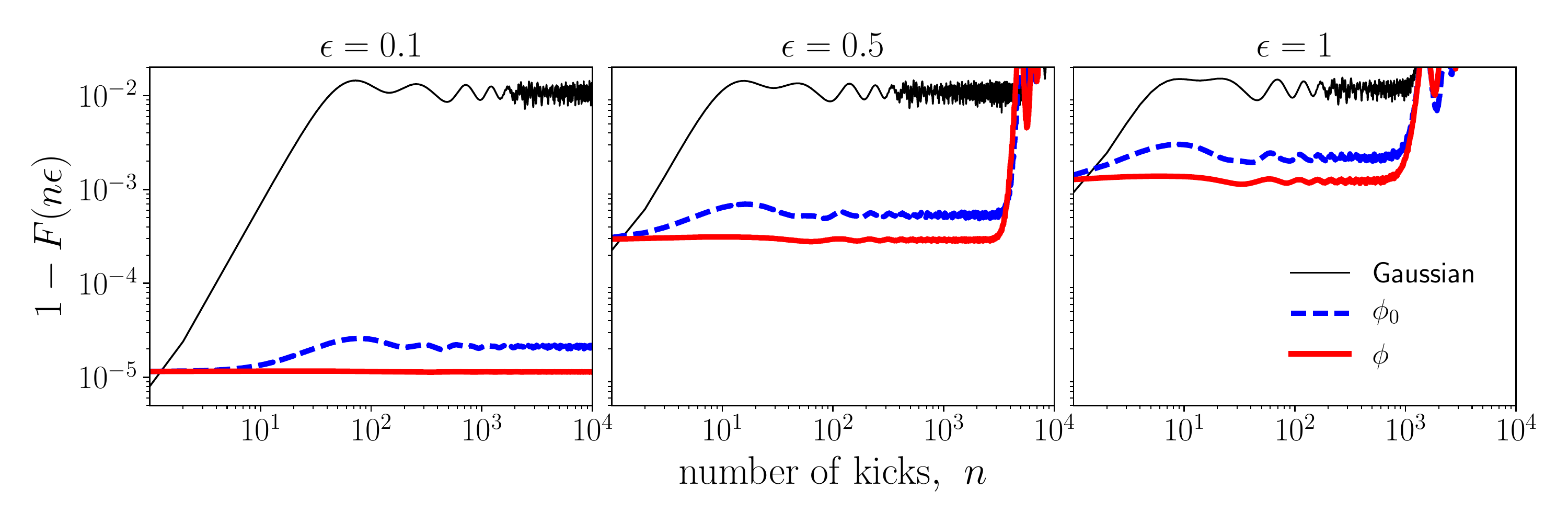}
	\caption{(Color online) Fidelity of stroboscopic solitons. Dependence of the fidelity error $1-F(n \epsilon)$ on the number of kicks for different kicking strength $\epsilon$,
	see Eq.~\eqref{God_parameter}. Three different initial states are considered: a Gaussian state (thin solid black curves), $\phi_0(x) = \tfrac{1}{2} \sech \frac{x}{2}$ (thick dashed blue curves), and $\phi(x)$ approximated by Eq.~(\ref{main_result}) (thick solid red curves).} 
	\label{fig:fidelity}
\end{figure*}

Now we consider the next order correction
\begin{equation}
	\phi_1 = \mu + i \nu \,,
\end{equation}
with $\mu(x)$ and $\nu(x)$ real-valued functions.  This yields (see Appendix~\ref{app:C})
\begin{align}
	&\mu'' + \left( -\tfrac{1}{4} + 6 \phi_0^2 \right) \mu = 0 \,, \label{eq-real_phi_1} \\
	&\nu'' + \left( -\tfrac{1}{4} + 2 \phi_0^2 \right) \nu = \phi_0^3 - 5 \phi_0^5 \,. \label{eq-imag_phi_1}
\end{align}
The general bounded solutions to Eqs.~(\ref{eq-real_phi_1}) and (\ref{eq-imag_phi_1}) are
\begin{align}
	&\mu = A \phi_0^2 \sinh \tfrac{x}{2} \,, \\
	&\nu = B \phi_0 + \tfrac{1}{2} \phi_0^3 \,,
\end{align}
where $A$ and $B$ are arbitrary real constants.
It turns out that the normalization condition, expanded up to $O(\epsilon^2)$, 
does not impose any constraints on the values of $A$ and $B$. Indeed, since $\phi_0$ is real, one has
\begin{equation}
	\int_{-\infty}^{+\infty} dx \, \real(\phi_0^* \phi_1) = \int_{-\infty}^{+\infty} dx \, \phi_0 \mu = 0 \qquad \forall\,A,B.
\end{equation}
A rigorous way of determining $A$ and $B$ would require one to proceed to a higher order in $\epsilon$, 
obtain the general expression for $\phi_2$ as a function of $A$ and $B$, and impose normalization up to order $O(\epsilon^3)$. 
The corresponding calculation however appears to be formidable.  Instead, we take $A = 0$, which ensures that $\phi(x)=\phi(-x)$, and $B = 0$, motivated by the observation that the quality of the soliton approximation, as quantified by fidelity (defined and discussed in detail below), is largely insensitive to the value of this parameter. 
Thus, we arrive at the following normalized approximation of the stationary stroboscopic soliton:
\begin{equation}
	\phi \simeq \left( 1 + \frac{\epsilon^2}{120} \right)^{-1/2} \left( \phi_0 + \frac{i \epsilon}{2} \phi_0^3 \right) \,.
\label{main_result}
\end{equation}
This constitutes the main analytical result of this paper.  
In view of Eq.~\eqref{eq-phi}, with $\alpha=1/8$, a one-parameter family of moving stroboscopic solitons 
is related to the stationary one via Eq.~(\ref{gauge}) with the dispersion relation
\begin{equation}
	\omega = \frac{1}{8} + \frac{v^2}{2} \,.
\end{equation}

\section{Numerical simulations}

In order to quantify the accuracy of Eq.~(\ref{main_result}), we investigate the problem numerically,
using the wave packet propagation algorithm Time-dependent Quantum Transport (TQT) \cite{krueckl2013}. 
Here both space and time are discretized, and the time-evolution operator is expanded in Krylov space during each time step, 
during which the Hamiltonian is assumed to be time-independent. We compute $\Psi(x,t)$ by taking a wave function $\Psi(x,0^+)$ and propagating it numerically according to Eq.~(\ref{NLSE-dimless}) with the $\delta$ function replaced by an appropriately normalized Gaussian, whose temporal width is short enough ($\lesssim \epsilon / 50$) in order to mimic an instantaneous perturbation.
We then quantify the overlap between the initial and the propagated wave functions computing the fidelity
\begin{equation}
	F(t) = \left| \int_{-\infty}^{+\infty} d x \, \Psi^*(x,0^+) \Psi(x,t) \right|^2 \,.
\label{fidelity}
\end{equation}
Initially, the fidelity equals unity, $F(0^+) = 1$, and generally decays in the course of time. However, if $\Psi(x,0^+)$ is chosen to 
coincide with the stroboscopic soliton wave function, Eq.~(\ref{main_result}), then $F(n \epsilon) \simeq 1$ for any integer $n$. Figure~\ref{fig:fidelity} shows the dependence of the quantity $1-F(n \epsilon)$ on $n$ for three different initial states $\Psi(x,0^+)$ and three different values of $\epsilon$, namely 0.1, 0.5 and 1; the smaller $1-F$, the closer is $\Psi(x,0^+)$ to a truly stroboscopic soliton wave function. Thin solid (black) curves correspond to $\Psi(x,0^+)$ being a Gaussian wave packet, thick dashed (blue) curves correspond to $\Psi(x,0^+) = \phi_0(x) = \tfrac{1}{2} \sech \tfrac{x}{2}$, and thick solid (red) curves correspond to $\Psi(x,0^+) = \phi(x)$, Eq.~(\ref{main_result}). The Gaussian wave packet is chosen to maximize its (fidelity) overlap with $\phi_0$, which fixes its initial spatial width to $\sigma \simeq 1.77$. The figure demonstrates how improving the $\epsilon$-order of the 
soliton approximation increases $F(n \epsilon)$. Indeed, the Gaussian initial state differs from the true stroboscopic soliton already at order $\epsilon^0$, which leads to a relatively low fidelity value. The error of taking $\Psi(x,0^+) = \phi_0(x)$ is of order $\epsilon^1$, and the corresponding fidelity error, $1-F$, is lower than that in the Gaussian case by roughly three orders of magnitude.  Finally, the error of approximating $\Psi(x,0^+)$ by Eq.~(\ref{main_result}) is of order $\epsilon^2$, and Fig.~\ref{fig:fidelity} clearly shows that the corresponding fidelity error becomes systematically lower in this case.  
Figure~\ref{fig:fidelity} also demonstrates that the fidelity $F(n \epsilon)$ decreases, though remaining at a high level, 
as $\epsilon$ increases; this is in accord with the fact that our analytical treatment is based on an expansion in $\epsilon \ll 1$.

We have also performed numerical analysis of the fidelity in the case when the initial wave packet moves with a nonzero velocity $v$
\footnote{Here, the fidelity definition, Eq.~(\ref{fidelity}), has to be modified in an obvious way in order to account for the wave packet drift in position space.}. 
We have found however no noticeable change compared to the stationary scenario: the fidelity curves corresponding to the $v \not= 0$ case appear to be almost indistinguishable from the ones shown in Fig.~\ref{fig:fidelity}. This observation reflects the fact that the problem involving a moving stroboscopic soliton can be mapped onto the stationary problem via the gauge transformation, Eq.~(\ref{gauge}).

\begin{figure}[h!]
	\centering
	\includegraphics[width=0.5\textwidth]{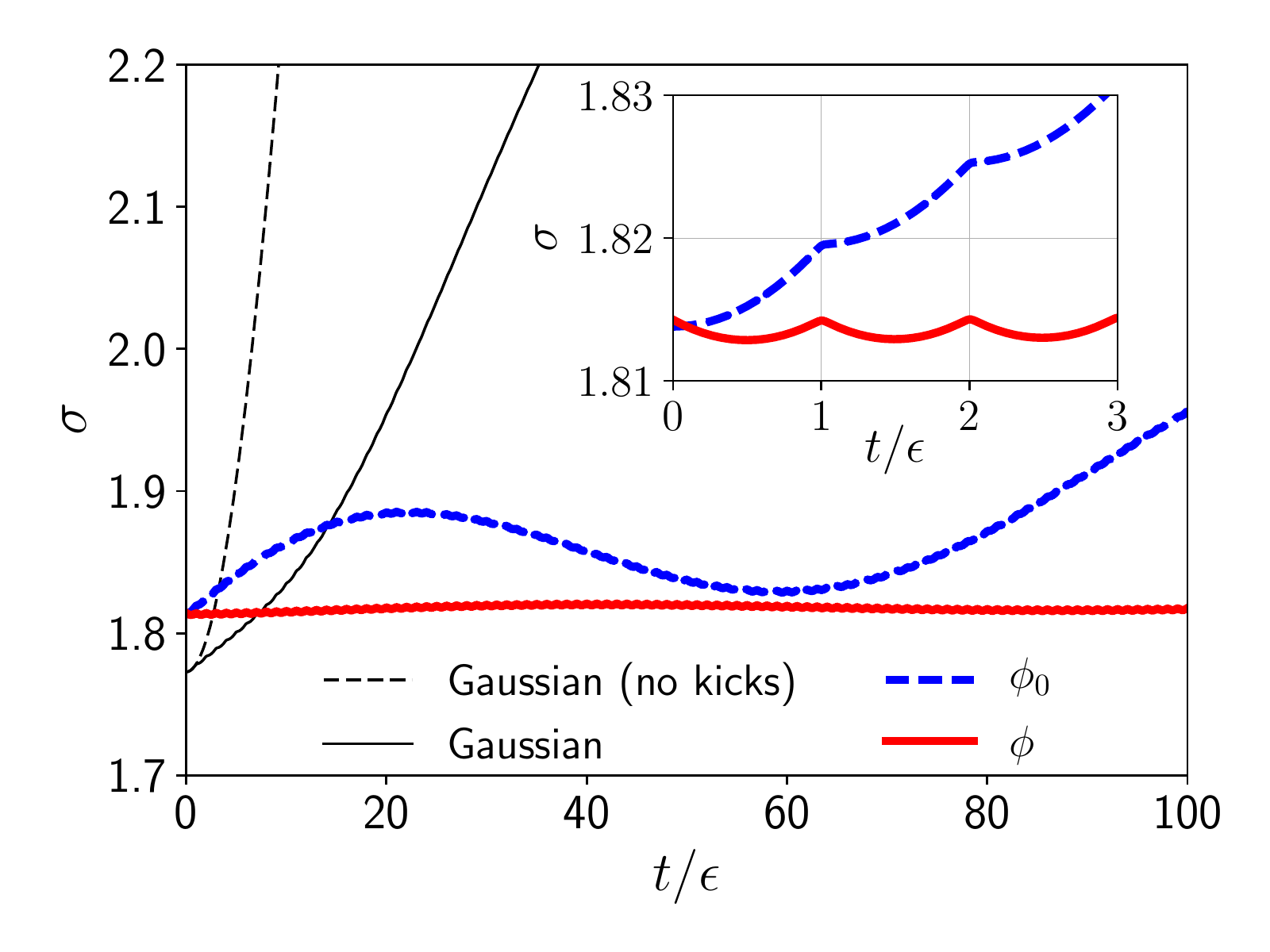}
	\caption{(Color online) Spatial width of $\Psi(x,t)$ as a function of $t$ for $\epsilon = 0.5$ and three different initial states: a Gaussian state (thin solid black curve), $\phi_0(x) = \tfrac{1}{2} \sech \frac{x}{2}$ (thick dashed blue curve), and $\phi(x)$ from Eq.~(\ref{main_result}) (thick solid red curve). The thin dashed black curve corresponds to the free-space evolution of the Gaussian wave packet. Inset shows short time behavior.} 
	\label{fig:width}
\end{figure}

The time dependence of the wave packet width 
\begin{equation}
	\sigma(t) = \sqrt{\langle x^2 \rangle_t - \langle x \rangle_t^2} \,,
\end{equation}
(where the average $\langle \cdot \rangle_t$ is defined with respect to the probability density $|\Psi(x,t)|^2$) is shown in Fig.~\ref{fig:width} for $\epsilon = 0.5$ and the three initial wave packets $\Psi(x,0^+)$ used in Fig.~\ref{fig:fidelity}; the color codes are the same in both figures. In addition to the three curves corresponding to the initial wave functions, $\sigma(t)$ is also shown for the case of a Gaussian wave packet evolving in the absence of any kicks, i.e. in free space (thin dashed black curve). We see that the introduction of nonlinear kicks significantly slows down the dispersive spreading of the Gaussian wave packet. The spreading is reduced further as one increases the $\epsilon$-order of accuracy of the stroboscopic soliton approximation. In particular, the average spreading of the wave packet initially given by Eq.~(\ref{main_result}) is almost entirely arrested. The inset in Fig.~\ref{fig:width} shows the function $\sigma(t)$ for $0 < t < 3 \epsilon$,
i.e. during the first three kicks. The width of the wave packet evolving starting from the real-valued function $\phi_0$ (thick dashed blue curve) increases monotonically until the first kick; the kick then changes the phase of the wave packet and, consequently, the rate of its spreading. The situation is qualitatively different in the case of the wave packet evolving from $\phi$, as given by Eq.~(\ref{main_result}) (thick solid red curve). Here, the wave function first focuses and then defocuses during the free flight in such a way that the width of $\Psi(x,\epsilon^-)$ is almost the same as that of $\Psi(x,0^+)$. This process then repeats itself from one kick to the next one, effectively giving rise to a localized matter wave with oscillating width.

Stroboscopically-dispersionless wave packets, proposed in this paper, could be realized in state-of-the-art atom-optics experiments involving ultracold atomic gases with temporally controlled scattering length \cite{Clark2015,nguyen2017,everitt2017,arunkumar2018}. 
The stroboscopic soliton-like states do not require any further stabilization potentials 
(e.g.~through background optical lattices~\cite{matuzewski2005}) and exist independent of the velocity.
This flexibility and robustness considerably simplify their possible realization.
In order to facilitate such a realization, we make an estimate of $\epsilon$, given by Eq.~(\ref{God_parameter}), that corresponds to a typical experimental setup. The dimensional kicking strength $\lambda$ is approximately equal to $2 N \hbar a_s \Delta t / (m a_{\perp}^2)$, where $N$ is the number of atoms, $\Delta t$ is the nonlinear kick duration, $a_s$ is the scattering length, and $a_\perp$ is the linear length scale of the potential confining the atomic motion in the transverse direction \cite{Salasnich2002}. Taking $N = 10^5$, $m = 7.016\,$u ($^7$Li atom), $a_s = 10\,n$m, $a_\perp = 10\,\mu$m, $\Delta t = 10\,\mu$s, as well as the duration between adjacent kicks $T = 5\,$ms, we obtain $\epsilon \simeq 0.072$. As confirmed by our 
numerical analysis, see Fig.~\ref{fig:fidelity}(a), this value of $\epsilon$ lies well within the range of validity of Eq.~(\ref{main_result}), implying stable solitonic dynamics of ultracold atoms on scales of seconds.

\section{Conclusion}

In conclusion, we have shown that dispersive spreading of a quantum wave packet 
can be stroboscopically undone by periodically kicking the nonlinear (interaction) term of the NLSE.  
The problem depends on a single parameter $\epsilon$, Eq.~\eqref{God_parameter}, independent of the wave packet velocity,
which translates into a wide range of physical regimes.  Moreover, our analytical solution was numerically shown to be robust 
beyond the perturbative $\epsilon\ll 1$ regime.  Coupled with the simplicity of the proposed protocol, this shows that
the realization of such a soliton-like object in atom optics experiments is well within the reach of current capabilities
\cite{Clark2015,nguyen2017,everitt2017,arunkumar2018}.

From a broader perspective, while engineering stable soliton-like waves via suitable periodic modulation 
of nonlinearity and dispersion is used in fiber optics
\footnote{For an overview see \cite{TBFlasers}.  
This type of optical pulses are typically obtained as solutions of the complex Ginzburg-Landau 
equation of breather type, that is their initial shape is periodically restored as the pulse propagates along the fibre, 
see e.g. \cite{Biondini,BBSTlocalized}.},
nonlinearity management in the context of quantum mechanical phenomena is incomparably less developed.
Recent proposals connected it with the physics of chaotic behavior \cite{zhao2016} and echoes \cite{reck2018}, 
and a powerful motivation for its study came very recently in the form of time crystals \cite{wilczek2012, sacha2018}.
In such a context it would be interesting to establish (or rule out) the existence, form and stability of (stroboscopic)
soliton-like solutions to nonlinear integral equations like \eqref{eq-phi-lin_op}, their possible connection
with recently proposed exotic states \cite{giergiel2018}, and their generalization to higher-dimensional systems.

\begin{acknowledgements}
	The authors thank Ilya Arakelyan and Gino Biondini for useful discussions, and Viktor Kr\"uckl for providing the TQT algorithm \cite{krueckl2013}. A.G. acknowledges the support of EPSRC Grant No.~EP/K024116/1, and thanks the Vielberth Foundation for supporting a stay at the University of Regensburg. P.R. thanks the Deutsche Forschungsgemeinschaft for financial support through RTG 1570.
\end{acknowledgements}

\appendix

\section{Derivation of the evolution operator}
\label{app:A}

Here we derive the evolution operator $U^{\epsilon}$ that propagates $\Psi(x,0)$ into $\Psi(x,\epsilon^+)$ in accordance with
\begin{equation*}
	i \frac{\partial \Psi}{\partial t} = -\frac{1}{2} \frac{\partial^2 \Psi}{\partial x^2} - \epsilon \delta(t - \epsilon) |\Psi|^2 \Psi \,.
\end{equation*}
To this end, we first construct the operator $\widetilde{U}_{\eta}^{\epsilon}$ that describes the evolution of $\Psi(x,0)$ into $\Psi(x,\epsilon+\eta)$ as governed by
\begin{equation*}
	i \frac{\partial \Psi}{\partial t} = -\frac{1}{2} \frac{\partial^2 \Psi}{\partial x^2} - \frac{\epsilon}{\eta} \big[ \theta(t-\epsilon) - \theta(t-\epsilon-\eta) \big]  |\Psi|^2 \Psi \,,
\end{equation*}
where $\theta(\tau)$ is the Heaviside step function. Then, $U^{\epsilon}$ can be found as
\begin{equation*}
	U^{\epsilon} = \lim_{\eta \rightarrow 0} \widetilde{U}_{\eta}^{\epsilon} \,.
\end{equation*}
Here, $\widetilde{U}_{\eta}^{\epsilon}$ satisfies the composition property
\begin{equation*}
	\widetilde{U}_{\eta}^{\epsilon} = \widetilde{K}_{\eta}^{\epsilon} U_0^{\epsilon} \,,
\end{equation*}
where $U_0^{\epsilon}$ is the free particle propagator, and $\widetilde{K}_{\eta}^{\epsilon}$ is the operator propagating $\Psi(x,\epsilon)$ into $\Psi(x,\epsilon+\eta)$ in accordance with
\begin{equation*}
	i \frac{\partial \Psi}{\partial t} = -\frac{1}{2} \frac{\partial^2 \Psi}{\partial x^2} - \frac{\epsilon}{\eta} |\Psi|^2 \Psi \,.
\end{equation*}
Similarly, we have
\begin{equation*}
	U^{\epsilon} = K^{\epsilon} U_0^{\epsilon} \quad \text{with} \quad K^{\epsilon} = \lim_{\eta \rightarrow 0} \widetilde{K}_{\eta}^{\epsilon} \,.
\end{equation*}
Making the substitution $\Psi = \sqrt{\rho} e^{i S}$, where $\rho(x,t)$ and $S(x,t)$ are real-valued functions, we rewrite the complex NLSE in the Hamilton-Jacobi form:
\begin{align*}
	&\frac{\partial \rho}{\partial t} = -\frac{\partial}{\partial x} \left( \rho \frac{\partial S}{\partial x} \right) \,, \\
	&\frac{\partial S}{\partial t} = \frac{1}{2 \sqrt{\rho}} \frac{\partial^2 \sqrt{\rho}}{\partial x^2} - \frac{1}{2} \left( \frac{\partial S}{\partial x} \right)^2 + \frac{\epsilon}{\eta} \rho \,,
\end{align*}
for $\epsilon < t < \epsilon + \eta$. Then, rescaling the time variable as $\tau = (t-\epsilon)/\eta$ and treating $\eta$ as a small parameter, we obtain
\begin{align*}
	&\frac{\partial \rho}{\partial \tau} = O(\eta) \,, \\
	&\frac{\partial S}{\partial \tau} = \epsilon \rho + O(\eta) \,,
\end{align*}
for $0 < \tau < 1$. Integrating this system of equations, we obtain $\rho \big|_{\tau = 1} = \rho \big|_{\tau = 0} + O(\eta)$ and $S \big|_{\tau = 1} = S \big|_{\tau = 0} + \epsilon \rho \big|_{\tau = 0} + O(\eta)$, or, in terms of the original time variable,
\begin{align*}
	&\rho(x,\epsilon+\eta) = \rho(x,\epsilon) + O(\eta) \,, \\
	&S(x,\epsilon+\eta) = S(x,\epsilon) + \epsilon \rho(x,\epsilon) + O(\eta) \,.
\end{align*}
Hence, for $\widetilde{K}_{\eta}^{\epsilon}$ we have
\begin{align*}
	K_{\eta}^{\epsilon} \Psi(x,\epsilon)
	&= \Psi(x,\epsilon+\eta) \\
	&= \sqrt{\rho(x,\epsilon)} e^{i S(x,\epsilon) + i \epsilon \rho(x,\epsilon)} + O(\eta) \\
	&= \Psi(x,\epsilon) e^{i \epsilon |\Psi(x,\epsilon)|^2} + O(\eta) \,.
\end{align*}
Finally, taking the limit $\eta \rightarrow 0$, we obtain the sought expression for the nonlinear kick operator:
\begin{equation*}
	K^{\epsilon} \Psi = \Psi e^{i \epsilon |\Psi|^2} \,.
\end{equation*}

\section{Derivation of Eq.~(\ref{eq-phi})}
\label{app:B}

Let us consider the equation describing moving stroboscopic solitons,
\begin{equation*}
	U^{\epsilon} \psi(x) = e^{i \omega \epsilon} \psi(x - v \epsilon) \,.
\end{equation*}
and make the substitution
\begin{equation*}
	\psi(x) = e^{i v x} \phi(x) \,.
\end{equation*}
Then, we have
\begin{align*}
	U^{\epsilon} \psi(x)
	&= K^{\epsilon} U_0^{\epsilon} e^{i v x} \phi(x) \\
	&= K^{\epsilon} \sqrt{\frac{1}{2 \pi i \epsilon}} \int_{-\infty}^{+\infty} dy \, e^{i (x-y)^2 / 2 \epsilon} e^{i v y} \phi(y) \\
	&= K^{\epsilon} e^{i v x - i v^2 \epsilon / 2} \sqrt{\frac{1}{2 \pi i \epsilon}} \int_{-\infty}^{+\infty} dy \, e^{i [ (x - v \epsilon) - y ]^2 / 2 \epsilon} \phi(y) \\
	&= K^{\epsilon} e^{i v x - i v^2 \epsilon / 2} U_0^{\epsilon} \phi(x - v \epsilon) \\
	&= e^{i v x - i v^2 \epsilon / 2} K^{\epsilon} U_0^{\epsilon} \phi(x - v \epsilon) \\
	&= e^{i v x - i v^2 \epsilon / 2} U^{\epsilon} \phi(x - v \epsilon) \,,
\end{align*}
and
\begin{align*}
	e^{i \omega \epsilon} \psi(x - v \epsilon)
	&= e^{i \omega \epsilon}  e^{i v (x - v \epsilon)} \phi(x - v \epsilon) \\
	&= e^{i v x + i (\omega - v^2) \epsilon} \phi(x - v \epsilon) \,.
\end{align*}
Equating the obtained expressions for $U^{\epsilon} \psi(x)$ and $e^{i \omega \epsilon} \psi(x - v \epsilon)$, and making the coordinate transformation to the moving reference frame, $z = x - v \epsilon$, we find the desired equation:
\begin{equation*}
	U^{\epsilon} \phi(z) = e^{i (\omega - v^2 / 2) \epsilon} \phi(z) \,.
\end{equation*}

\onecolumngrid

\section{Derivation of Eqs.~(\ref{phi_0_eq}), (\ref{eq-real_phi_1}) and (\ref{eq-imag_phi_1})}
\label{app:C}

Here, we look for an asymptotic form of the stroboscopic soliton equation,
\begin{equation*}
	U_0^{\epsilon} \phi = \phi e^{ i \epsilon \left( \alpha - |\phi|^2 \right) } \,,
\end{equation*}
by expanding the wave function into a power series in $\epsilon$,
\begin{equation*}
	\phi = \phi_0 + \epsilon \phi_1 + \epsilon^2 \phi_2 + O(\epsilon^3) \,.
\end{equation*}
For the left-hand side of the equation, we have
\begin{align*}
	U_0^{\epsilon} \phi
	&= \phi + \epsilon \tfrac{i}{2} \phi'' - \epsilon^2 \tfrac{1}{8} \phi^{(4)} + O(\epsilon^3) \\
	&= \phi_0 + \epsilon \left( \tfrac{i}{2} \phi''_0 +  \phi_1 \right) + \epsilon^2 \left( -\tfrac{1}{8} \phi''''_0 + \tfrac{i}{2} \phi''_1 + \phi_2 \right) + O(\epsilon^3) \,.
\end{align*}
The right-hand side is expanded as follows. Since
\begin{equation*}
	|\phi|^2 = |\phi_0|^2 + 2 \epsilon \real(\phi_0^* \phi_1) + O(\epsilon^2) \,,
\end{equation*}
we have
\begin{align*}
	\phi e^{ i \epsilon \left( \alpha - |\phi|^2 \right) }
	&= \left[ \phi_0 + \epsilon \phi_1 + \epsilon^2 \phi_2 + O(\epsilon^3) \right] \left[ 1 + i \epsilon (\alpha - |\phi|^2) - \tfrac{1}{2} \epsilon^2 (\alpha - |\phi|^2)^2 + O(\epsilon^3) \right] \\
	&= \phi_0 + \epsilon \left[ i (\alpha - |\phi_0|^2) \phi_0 + \phi_1 \right] + \epsilon^2 \left[ -\tfrac{1}{2} (\alpha - |\phi_0|^2)^2 \phi_0 - 2 i \phi_0 \real(\phi_0^* \phi_1) + i (\alpha - |\phi_0|^2) \phi_1 + \phi_2 \right] + O(\epsilon^3) \,.
\end{align*}
Now, we compare the obtained expressions for the left- and right-hand sides order by order in $\epsilon$. At order $\epsilon^0$, we get a trivial identity. At order $\epsilon^1$, we obtain
\begin{equation*}
	-\tfrac{1}{2} \phi''_0 + \left( \alpha - |\phi_0|^2 \right) \phi_0 = 0 \,.
\end{equation*}
At order $\epsilon^2$, we find the equation
\begin{equation*}
	-\tfrac{1}{2} \phi''_1 + \left( \alpha - |\phi_0|^2 \right) \phi_1 - 2 \phi_0 \real(\phi_0^* \phi_1) = \tfrac{i}{8} \phi''''_0 - \tfrac{i}{2} (\alpha - |\phi_0|^2)^2 \phi_0 \,.
\end{equation*}
Upon substituting $\phi_1 = \mu + i \nu$, and taking into account that $\phi_0 = \tfrac{1}{2} \sech \tfrac{x}{2}$ and $\alpha = \tfrac{1}{8}$, the last equation splits into the two desired equations for $\mu$ and $\nu$.

\bigskip

\twocolumngrid


%


\begin{thebibliography}{31}%
\makeatletter
\providecommand \@ifxundefined [1]{%
 \@ifx{#1\undefined}
}%
\providecommand \@ifnum [1]{%
 \ifnum #1\expandafter \@firstoftwo
 \else \expandafter \@secondoftwo
 \fi
}%
\providecommand \@ifx [1]{%
 \ifx #1\expandafter \@firstoftwo
 \else \expandafter \@secondoftwo
 \fi
}%
\providecommand \natexlab [1]{#1}%
\providecommand \enquote  [1]{``#1''}%
\providecommand \bibnamefont  [1]{#1}%
\providecommand \bibfnamefont [1]{#1}%
\providecommand \citenamefont [1]{#1}%
\providecommand \href@noop [0]{\@secondoftwo}%
\providecommand \href [0]{\begingroup \@sanitize@url \@href}%
\providecommand \@href[1]{\@@startlink{#1}\@@href}%
\providecommand \@@href[1]{\endgroup#1\@@endlink}%
\providecommand \@sanitize@url [0]{\catcode `\\12\catcode `\$12\catcode
  `\&12\catcode `\#12\catcode `\^12\catcode `\_12\catcode `\%12\relax}%
\providecommand \@@startlink[1]{}%
\providecommand \@@endlink[0]{}%
\providecommand \url  [0]{\begingroup\@sanitize@url \@url }%
\providecommand \@url [1]{\endgroup\@href {#1}{\urlprefix }}%
\providecommand \urlprefix  [0]{URL }%
\providecommand \Eprint [0]{\href }%
\providecommand \doibase [0]{http://dx.doi.org/}%
\providecommand \selectlanguage [0]{\@gobble}%
\providecommand \bibinfo  [0]{\@secondoftwo}%
\providecommand \bibfield  [0]{\@secondoftwo}%
\providecommand \translation [1]{[#1]}%
\providecommand \BibitemOpen [0]{}%
\providecommand \bibitemStop [0]{}%
\providecommand \bibitemNoStop [0]{.\EOS\space}%
\providecommand \EOS [0]{\spacefactor3000\relax}%
\providecommand \BibitemShut  [1]{\csname bibitem#1\endcsname}%
\let\auto@bib@innerbib\@empty
\bibitem [{\citenamefont {Berry}\ and\ \citenamefont
  {Balazs}(1979)}]{BB79Nonspreading}%
  \BibitemOpen
  \bibfield  {author} {\bibinfo {author} {\bibfnamefont {M.~V.}\ \bibnamefont
  {Berry}}\ and\ \bibinfo {author} {\bibfnamefont {N.~L.}\ \bibnamefont
  {Balazs}},\ }\bibfield  {title} {\enquote {\bibinfo {title} {Nonspreading
  wave packets},}\ }\href@noop {} {\bibfield  {journal} {\bibinfo  {journal}
  {Am. J. Phys.}\ }\textbf {\bibinfo {volume} {47}},\ \bibinfo {pages} {264}
  (\bibinfo {year} {1979})}\BibitemShut {NoStop}%
\bibitem [{\citenamefont {Greenberger}(1980)}]{Gre80Comment}%
  \BibitemOpen
  \bibfield  {author} {\bibinfo {author} {\bibfnamefont {D.~M.}\ \bibnamefont
  {Greenberger}},\ }\bibfield  {title} {\enquote {\bibinfo {title} {Comment on
  ``{N}onspreading wave packets''},}\ }\href@noop {} {\bibfield  {journal}
  {\bibinfo  {journal} {Am. J. Phys.}\ }\textbf {\bibinfo {volume} {48}},\
  \bibinfo {pages} {256} (\bibinfo {year} {1980})}\BibitemShut {NoStop}%
\bibitem [{\citenamefont {Unnikrishnan}\ and\ \citenamefont
  {Rau}(1996)}]{UR96Uniqueness}%
  \BibitemOpen
  \bibfield  {author} {\bibinfo {author} {\bibfnamefont {K.}~\bibnamefont
  {Unnikrishnan}}\ and\ \bibinfo {author} {\bibfnamefont {A.~R.~P.}\
  \bibnamefont {Rau}},\ }\bibfield  {title} {\enquote {\bibinfo {title}
  {Uniqueness of the {A}iry packet in quantum mechanics},}\ }\href@noop {}
  {\bibfield  {journal} {\bibinfo  {journal} {Am. J. Phys.}\ }\textbf {\bibinfo
  {volume} {64}},\ \bibinfo {pages} {1034} (\bibinfo {year}
  {1996})}\BibitemShut {NoStop}%
\bibitem [{\citenamefont {Schr{\"o}dinger}(1926)}]{Sch26stetige}%
  \BibitemOpen
  \bibfield  {author} {\bibinfo {author} {\bibfnamefont {E.}~\bibnamefont
  {Schr{\"o}dinger}},\ }\bibfield  {title} {\enquote {\bibinfo {title} {Der
  stetige {\"u}bergang von der {M}ikro- zur {M}akromechanik},}\ }\href@noop {}
  {\bibfield  {journal} {\bibinfo  {journal} {Naturwissenschaften}\ }\textbf
  {\bibinfo {volume} {14}},\ \bibinfo {pages} {664} (\bibinfo {year}
  {1926})}\BibitemShut {NoStop}%
\bibitem [{\citenamefont {Zhang}\ \emph {et~al.}(1990)\citenamefont {Zhang},
  \citenamefont {Feng},\ and\ \citenamefont {Gilmore}}]{ZFG90Coherent}%
  \BibitemOpen
  \bibfield  {author} {\bibinfo {author} {\bibfnamefont {W.-M.}\ \bibnamefont
  {Zhang}}, \bibinfo {author} {\bibfnamefont {D.~H.}\ \bibnamefont {Feng}}, \
  and\ \bibinfo {author} {\bibfnamefont {R.}~\bibnamefont {Gilmore}},\
  }\bibfield  {title} {\enquote {\bibinfo {title} {Coherent states: {T}heory
  and some applications},}\ }\href@noop {} {\bibfield  {journal} {\bibinfo
  {journal} {Rev. Mod. Phys.}\ }\textbf {\bibinfo {volume} {62}},\ \bibinfo
  {pages} {867} (\bibinfo {year} {1990})}\BibitemShut {NoStop}%
\bibitem [{\citenamefont {Dodonov}\ and\ \citenamefont
  {Man'ko}(2003)}]{DM03Theory}%
  \BibitemOpen
  \bibinfo {editor} {\bibfnamefont {V.~V.}\ \bibnamefont {Dodonov}}\ and\
  \bibinfo {editor} {\bibfnamefont {V.~I.}\ \bibnamefont {Man'ko}},\ eds.,\
  \href@noop {} {\emph {\bibinfo {title} {Theory of Nonclassical States of
  Light}}}\ (\bibinfo  {publisher} {Taylor \& {F}rancis, {L}ondon and {N}ew
  {Y}ork},\ \bibinfo {year} {2003})\BibitemShut {NoStop}%
\bibitem [{\citenamefont {Gazeau}(2009)}]{Gaz09Coherent}%
  \BibitemOpen
  \bibfield  {author} {\bibinfo {author} {\bibfnamefont {J.-P.}\ \bibnamefont
  {Gazeau}},\ }\href@noop {} {\emph {\bibinfo {title} {Coherent States in
  Quantum Physics}}}\ (\bibinfo  {publisher} {Wiley-{V}{C}{H}, Berlin},\
  \bibinfo {year} {2009})\BibitemShut {NoStop}%
\bibitem [{\citenamefont {Dodonov}(2018)}]{DodCoherent}%
  \BibitemOpen
  \bibfield  {author} {\bibinfo {author} {\bibfnamefont {V.~V.}\ \bibnamefont
  {Dodonov}},\ }\bibfield  {title} {\enquote {\bibinfo {title} {Coherent states
  and their generalizations for a charged particle in a magnetic field},}\ }in\
  \href@noop {} {\emph {\bibinfo {booktitle} {Springer Proceedings in Physics
  205}}},\ \bibinfo {editor} {edited by\ \bibinfo {editor} {\bibfnamefont
  {J.-P.}\ \bibnamefont {Antoine}}, \bibinfo {editor} {\bibfnamefont
  {F.}~\bibnamefont {Bagarello}}, \ and\ \bibinfo {editor} {\bibfnamefont
  {J.-P.}\ \bibnamefont {Gazeau}}}\ (\bibinfo  {publisher} {Springer},\
  \bibinfo {year} {2018})\BibitemShut {NoStop}%
\bibitem [{\citenamefont {Buchleitner}\ \emph {et~al.}(2002)\citenamefont
  {Buchleitner}, \citenamefont {Delande},\ and\ \citenamefont
  {Zakrzewski}}]{BDZ02Non-dispersive}%
  \BibitemOpen
  \bibfield  {author} {\bibinfo {author} {\bibfnamefont {A.}~\bibnamefont
  {Buchleitner}}, \bibinfo {author} {\bibfnamefont {D.}~\bibnamefont
  {Delande}}, \ and\ \bibinfo {author} {\bibfnamefont {J.}~\bibnamefont
  {Zakrzewski}},\ }\bibfield  {title} {\enquote {\bibinfo {title}
  {Non-dispersive wave packets in periodically driven quantum systems},}\
  }\href@noop {} {\bibfield  {journal} {\bibinfo  {journal} {Phys. Rep.}\
  }\textbf {\bibinfo {volume} {368}},\ \bibinfo {pages} {409} (\bibinfo {year}
  {2002})}\BibitemShut {NoStop}%
\bibitem [{\citenamefont {Kivshar}\ and\ \citenamefont
  {Malomed}(1989)}]{KM89Dynamics}%
  \BibitemOpen
  \bibfield  {author} {\bibinfo {author} {\bibfnamefont {Y.~S.}\ \bibnamefont
  {Kivshar}}\ and\ \bibinfo {author} {\bibfnamefont {B.~A.}\ \bibnamefont
  {Malomed}},\ }\bibfield  {title} {\enquote {\bibinfo {title} {Dynamics of
  solitons in nearly integrable systems},}\ }\href@noop {} {\bibfield
  {journal} {\bibinfo  {journal} {Rev. Mod. Phys.}\ }\textbf {\bibinfo {volume}
  {61}},\ \bibinfo {pages} {763} (\bibinfo {year} {1989})}\BibitemShut
  {NoStop}%
\bibitem [{\citenamefont {Ablowitz}\ \emph {et~al.}(2004)\citenamefont
  {Ablowitz}, \citenamefont {Prinari},\ and\ \citenamefont
  {Trubatch}}]{APTNLS}%
  \BibitemOpen
  \bibfield  {author} {\bibinfo {author} {\bibfnamefont {M.~J.}\ \bibnamefont
  {Ablowitz}}, \bibinfo {author} {\bibfnamefont {B.}~\bibnamefont {Prinari}}, \
  and\ \bibinfo {author} {\bibfnamefont {A.~D.}\ \bibnamefont {Trubatch}},\
  }\href@noop {} {\emph {\bibinfo {title} {Discrete and Continuous Nonlinear
  Schr\"odinger Systems}}}\ (\bibinfo  {publisher} {Cambridge University
  Press},\ \bibinfo {year} {2004})\BibitemShut {NoStop}%
\bibitem [{\citenamefont {Kevrekidis}\ \emph {et~al.}(2003)\citenamefont
  {Kevrekidis}, \citenamefont {Theocharis}, \citenamefont {Frantzeskakis},\
  and\ \citenamefont {Malomed}}]{kevrekidis2003}%
  \BibitemOpen
  \bibfield  {author} {\bibinfo {author} {\bibfnamefont {P.~G.}\ \bibnamefont
  {Kevrekidis}}, \bibinfo {author} {\bibfnamefont {G.}~\bibnamefont
  {Theocharis}}, \bibinfo {author} {\bibfnamefont {D.~J.}\ \bibnamefont
  {Frantzeskakis}}, \ and\ \bibinfo {author} {\bibfnamefont {Boris~A.}\
  \bibnamefont {Malomed}},\ }\bibfield  {title} {\enquote {\bibinfo {title}
  {Feshbach resonance management for bose-einstein condensates},}\ }\href
  {https://journals.aps.org/prl/abstract/10.1103/PhysRevLett.90.230401}
  {\bibfield  {journal} {\bibinfo  {journal} {Phys. Rev. Lett.}\ }\textbf
  {\bibinfo {volume} {90}},\ \bibinfo {pages} {230401} (\bibinfo {year}
  {2003})}\BibitemShut {NoStop}%
\bibitem [{Note1()}]{Note1}%
  \BibitemOpen
  \bibinfo {note} {According to the botanics of Ref.~\cite {bukov2015}, such
  systems belong to the ``Kapitza class''}\BibitemShut {NoStop}%
\bibitem [{\citenamefont {Clark}\ \emph {et~al.}(2015)\citenamefont {Clark},
  \citenamefont {Ha}, \citenamefont {Xu},\ and\ \citenamefont
  {Chin}}]{Clark2015}%
  \BibitemOpen
  \bibfield  {author} {\bibinfo {author} {\bibfnamefont {L.~W.}\ \bibnamefont
  {Clark}}, \bibinfo {author} {\bibfnamefont {L.}~\bibnamefont {Ha}}, \bibinfo
  {author} {\bibfnamefont {C.}~\bibnamefont {Xu}}, \ and\ \bibinfo {author}
  {\bibfnamefont {C.}~\bibnamefont {Chin}},\ }\bibfield  {title} {\enquote
  {\bibinfo {title} {Quantum dynamics with spatiotemporal control of
  interactions in a stable {B}ose-{E}instein condensate},}\ }\href@noop {}
  {\bibfield  {journal} {\bibinfo  {journal} {Phys. Rev. Lett.}\ }\textbf
  {\bibinfo {volume} {115}},\ \bibinfo {pages} {155301} (\bibinfo {year}
  {2015})}\BibitemShut {NoStop}%
\bibitem [{\citenamefont {Nguyen}\ \emph {et~al.}(2017)\citenamefont {Nguyen},
  \citenamefont {Luo},\ and\ \citenamefont {Hulet}}]{nguyen2017}%
  \BibitemOpen
  \bibfield  {author} {\bibinfo {author} {\bibfnamefont {J.~H.~V.}\
  \bibnamefont {Nguyen}}, \bibinfo {author} {\bibfnamefont {D.}~\bibnamefont
  {Luo}}, \ and\ \bibinfo {author} {\bibfnamefont {R.~G.}\ \bibnamefont
  {Hulet}},\ }\bibfield  {title} {\enquote {\bibinfo {title} {Formation of
  matter-wave soliton trains by modulational instability},}\ }\href
  {http://science.sciencemag.org/content/356/6336/422} {\bibfield  {journal}
  {\bibinfo  {journal} {Science}\ }\textbf {\bibinfo {volume} {356}},\ \bibinfo
  {pages} {422} (\bibinfo {year} {2017})}\BibitemShut {NoStop}%
\bibitem [{\citenamefont {Everitt}\ \emph {et~al.}(2017)\citenamefont
  {Everitt}, \citenamefont {Sooriyabandara}, \citenamefont {Guasoni},
  \citenamefont {Wigley}, \citenamefont {Wei}, \citenamefont {McDonald},
  \citenamefont {Hardman}, \citenamefont {Manju}, \citenamefont {Close},
  \citenamefont {Kuhn}, \citenamefont {Szigeti}, \citenamefont {Kivshar},\ and\
  \citenamefont {Robins}}]{everitt2017}%
  \BibitemOpen
  \bibfield  {author} {\bibinfo {author} {\bibfnamefont {P.~J.}\ \bibnamefont
  {Everitt}}, \bibinfo {author} {\bibfnamefont {M.~A.}\ \bibnamefont
  {Sooriyabandara}}, \bibinfo {author} {\bibfnamefont {M.}~\bibnamefont
  {Guasoni}}, \bibinfo {author} {\bibfnamefont {P.~B.}\ \bibnamefont {Wigley}},
  \bibinfo {author} {\bibfnamefont {C.~H.}\ \bibnamefont {Wei}}, \bibinfo
  {author} {\bibfnamefont {G.~D.}\ \bibnamefont {McDonald}}, \bibinfo {author}
  {\bibfnamefont {K.~S.}\ \bibnamefont {Hardman}}, \bibinfo {author}
  {\bibfnamefont {P.}~\bibnamefont {Manju}}, \bibinfo {author} {\bibfnamefont
  {J.~D.}\ \bibnamefont {Close}}, \bibinfo {author} {\bibfnamefont {C.~C.~N.}\
  \bibnamefont {Kuhn}}, \bibinfo {author} {\bibfnamefont {S.~S.}\ \bibnamefont
  {Szigeti}}, \bibinfo {author} {\bibfnamefont {Y.~S.}\ \bibnamefont
  {Kivshar}}, \ and\ \bibinfo {author} {\bibfnamefont {N.~P.}\ \bibnamefont
  {Robins}},\ }\bibfield  {title} {\enquote {\bibinfo {title} {Observation of a
  modulational instability in {B}ose-{E}instein condensates},}\ }\href@noop {}
  {\bibfield  {journal} {\bibinfo  {journal} {Phys. Rev. A}\ }\textbf {\bibinfo
  {volume} {96}},\ \bibinfo {pages} {041601(R)} (\bibinfo {year}
  {2017})}\BibitemShut {NoStop}%
\bibitem [{\citenamefont {Arunkumar}\ \emph {et~al.}(2018)\citenamefont
  {Arunkumar}, \citenamefont {Jagannathan},\ and\ \citenamefont
  {Thomas}}]{arunkumar2018}%
  \BibitemOpen
  \bibfield  {author} {\bibinfo {author} {\bibfnamefont {N.}~\bibnamefont
  {Arunkumar}}, \bibinfo {author} {\bibfnamefont {A.}~\bibnamefont
  {Jagannathan}}, \ and\ \bibinfo {author} {\bibfnamefont {J.~E.}\ \bibnamefont
  {Thomas}},\ }\bibfield  {title} {\enquote {\bibinfo {title} {Designer spatial
  control of interactions in ultracold gases},}\ }\href
  {https://arxiv.org/abs/1803.01920} {\bibfield  {journal} {\bibinfo  {journal}
  {arXiv}\ ,\ \bibinfo {pages} {1803.01920}} (\bibinfo {year}
  {2018})}\BibitemShut {NoStop}%
\bibitem [{\citenamefont {Kr\"{u}ckl}(2013)}]{krueckl2013}%
  \BibitemOpen
  \bibfield  {author} {\bibinfo {author} {\bibfnamefont {V.}~\bibnamefont
  {Kr\"{u}ckl}},\ }\bibfield  {title} {\enquote {\bibinfo {title} {{Wave
  packets in mesoscopic systems: From time-dependent dynamics to transport
  phenomena in graphene and topological insulators}},}\ }\href
  {http://epub.uni-regensburg.de/28081/} {\  (\bibinfo {year} {2013})},\
  \bibinfo {note} {the basic version of the algorithm is available at
  \href{http://www.krueckl.de/\#en/tqt.php}{TQT Home}
  {[}http://www.krueckl.de/\#en/tqt.php{]}.}\BibitemShut {Stop}%
\bibitem [{Note2()}]{Note2}%
  \BibitemOpen
  \bibinfo {note} {Here, the fidelity definition, Eq.~(\ref {fidelity}), has to
  be modified in an obvious way in order to account for the wave packet drift
  in position space.}\BibitemShut {Stop}%
\bibitem [{\citenamefont {Matuszewski}\ \emph {et~al.}(2005)\citenamefont
  {Matuszewski}, \citenamefont {Infeld}, \citenamefont {Malomed},\ and\
  \citenamefont {Trippenbach}}]{matuzewski2005}%
  \BibitemOpen
  \bibfield  {author} {\bibinfo {author} {\bibfnamefont {M.}~\bibnamefont
  {Matuszewski}}, \bibinfo {author} {\bibfnamefont {E.}~\bibnamefont {Infeld}},
  \bibinfo {author} {\bibfnamefont {B.~A.}\ \bibnamefont {Malomed}}, \ and\
  \bibinfo {author} {\bibfnamefont {M.}~\bibnamefont {Trippenbach}},\
  }\bibfield  {title} {\enquote {\bibinfo {title} {Fully three dimensional
  breather solitons can be created using feshbach resonances},}\ }\href@noop {}
  {\bibfield  {journal} {\bibinfo  {journal} {Phys. Rev. Lett.}\ }\textbf
  {\bibinfo {volume} {95}},\ \bibinfo {pages} {050403} (\bibinfo {year}
  {2005})}\BibitemShut {NoStop}%
\bibitem [{\citenamefont {Salasnich}\ \emph {et~al.}(2002)\citenamefont
  {Salasnich}, \citenamefont {Parola},\ and\ \citenamefont
  {Reatto}}]{Salasnich2002}%
  \BibitemOpen
  \bibfield  {author} {\bibinfo {author} {\bibfnamefont {L.}~\bibnamefont
  {Salasnich}}, \bibinfo {author} {\bibfnamefont {A.}~\bibnamefont {Parola}}, \
  and\ \bibinfo {author} {\bibfnamefont {L.}~\bibnamefont {Reatto}},\
  }\bibfield  {title} {\enquote {\bibinfo {title} {Effective wave equations for
  the dynamics of cigar-shaped and disk-shaped {B}ose condensate},}\
  }\href@noop {} {\bibfield  {journal} {\bibinfo  {journal} {Phys. Rev. A}\
  }\textbf {\bibinfo {volume} {65}},\ \bibinfo {pages} {43614} (\bibinfo {year}
  {2002})}\BibitemShut {NoStop}%
\bibitem [{Note3()}]{Note3}%
  \BibitemOpen
  \bibinfo {note} {For an overview see \cite {TBFlasers}. This type of optical
  pulses are typically obtained as solutions of the complex Ginzburg-Landau
  equation of breather type, that is their initial shape is periodically
  restored as the pulse propagates along the fibre, see e.g. \cite
  {Biondini,BBSTlocalized}.}\BibitemShut {Stop}%
\bibitem [{\citenamefont {Zhao}\ \emph {et~al.}(2016)\citenamefont {Zhao},
  \citenamefont {Gong}, \citenamefont {Wang}, \citenamefont {Casati},
  \citenamefont {Liu},\ and\ \citenamefont {Fu}}]{zhao2016}%
  \BibitemOpen
  \bibfield  {author} {\bibinfo {author} {\bibfnamefont {W.-L.}\ \bibnamefont
  {Zhao}}, \bibinfo {author} {\bibfnamefont {J.}~\bibnamefont {Gong}}, \bibinfo
  {author} {\bibfnamefont {W.-G.}\ \bibnamefont {Wang}}, \bibinfo {author}
  {\bibfnamefont {G.}~\bibnamefont {Casati}}, \bibinfo {author} {\bibfnamefont
  {J.}~\bibnamefont {Liu}}, \ and\ \bibinfo {author} {\bibfnamefont {L.-B.}\
  \bibnamefont {Fu}},\ }\bibfield  {title} {\enquote {\bibinfo {title}
  {Exponential wave-packet spreading via self-interaction time modulation},}\
  }\href {https://journals.aps.org/pra/abstract/10.1103/PhysRevA.94.053631}
  {\bibfield  {journal} {\bibinfo  {journal} {Phys. Rev. A}\ }\textbf {\bibinfo
  {volume} {94}},\ \bibinfo {pages} {053631} (\bibinfo {year}
  {2016})}\BibitemShut {NoStop}%
\bibitem [{\citenamefont {Reck}\ \emph {et~al.}(2018)\citenamefont {Reck},
  \citenamefont {Gorini}, \citenamefont {Goussev}, \citenamefont {Krueckl},
  \citenamefont {Fink},\ and\ \citenamefont {Richter}}]{reck2018}%
  \BibitemOpen
  \bibfield  {author} {\bibinfo {author} {\bibfnamefont {P.}~\bibnamefont
  {Reck}}, \bibinfo {author} {\bibfnamefont {C.}~\bibnamefont {Gorini}},
  \bibinfo {author} {\bibfnamefont {A.}~\bibnamefont {Goussev}}, \bibinfo
  {author} {\bibfnamefont {V.}~\bibnamefont {Krueckl}}, \bibinfo {author}
  {\bibfnamefont {M.}~\bibnamefont {Fink}}, \ and\ \bibinfo {author}
  {\bibfnamefont {K.}~\bibnamefont {Richter}},\ }\bibfield  {title} {\enquote
  {\bibinfo {title} {Towards a quantum time mirror for non-relativistic wave
  packets},}\ }\href
  {http://iopscience.iop.org/article/10.1088/1367-2630/aaae98/meta} {\bibfield
  {journal} {\bibinfo  {journal} {New J. Phys.}\ }\textbf {\bibinfo {volume}
  {20}},\ \bibinfo {pages} {033013} (\bibinfo {year} {2018})}\BibitemShut
  {NoStop}%
\bibitem [{\citenamefont {Wilczek}(2012)}]{wilczek2012}%
  \BibitemOpen
  \bibfield  {author} {\bibinfo {author} {\bibfnamefont {F.}~\bibnamefont
  {Wilczek}},\ }\bibfield  {title} {\enquote {\bibinfo {title} {Quantum time
  crystals},}\ }\href
  {https://journals.aps.org/prl/abstract/10.1103/PhysRevLett.109.160401}
  {\bibfield  {journal} {\bibinfo  {journal} {Phys. Rev. Lett.}\ }\textbf
  {\bibinfo {volume} {109}},\ \bibinfo {pages} {160401} (\bibinfo {year}
  {2012})}\BibitemShut {NoStop}%
\bibitem [{\citenamefont {Sacha}\ and\ \citenamefont
  {Zakrzewski}(2018)}]{sacha2018}%
  \BibitemOpen
  \bibfield  {author} {\bibinfo {author} {\bibfnamefont {K.}~\bibnamefont
  {Sacha}}\ and\ \bibinfo {author} {\bibfnamefont {J.}~\bibnamefont
  {Zakrzewski}},\ }\bibfield  {title} {\enquote {\bibinfo {title} {Time
  crystals: a review},}\ }\href
  {http://iopscience.iop.org/article/10.1088/1361-6633/aa8b38/meta} {\bibfield
  {journal} {\bibinfo  {journal} {Rep. Prog. Phys.}\ }\textbf {\bibinfo
  {volume} {81}},\ \bibinfo {pages} {016401} (\bibinfo {year}
  {2018})}\BibitemShut {NoStop}%
\bibitem [{\citenamefont {Giergiel}\ \emph {et~al.}(2018)\citenamefont
  {Giergiel}, \citenamefont {Miroszewski},\ and\ \citenamefont
  {Sacha}}]{giergiel2018}%
  \BibitemOpen
  \bibfield  {author} {\bibinfo {author} {\bibfnamefont {K.}~\bibnamefont
  {Giergiel}}, \bibinfo {author} {\bibfnamefont {A.}~\bibnamefont
  {Miroszewski}}, \ and\ \bibinfo {author} {\bibfnamefont {K.}~\bibnamefont
  {Sacha}},\ }\bibfield  {title} {\enquote {\bibinfo {title} {Time crystal
  platform: From quasicrystal structures in time to systems with exotic
  interactions},}\ }\href
  {https://journals.aps.org/prl/abstract/10.1103/PhysRevLett.120.140401}
  {\bibfield  {journal} {\bibinfo  {journal} {Phys. Rev. Lett.}\ }\textbf
  {\bibinfo {volume} {120}},\ \bibinfo {pages} {140401} (\bibinfo {year}
  {2018})}\BibitemShut {NoStop}%
\bibitem [{\citenamefont {Bukov}\ \emph {et~al.}(2015)\citenamefont {Bukov},
  \citenamefont {D'Alessio},\ and\ \citenamefont {Polkovnikov}}]{bukov2015}%
  \BibitemOpen
  \bibfield  {author} {\bibinfo {author} {\bibfnamefont {M.}~\bibnamefont
  {Bukov}}, \bibinfo {author} {\bibfnamefont {L.}~\bibnamefont {D'Alessio}}, \
  and\ \bibinfo {author} {\bibfnamefont {A.}~\bibnamefont {Polkovnikov}},\
  }\bibfield  {title} {\enquote {\bibinfo {title} {Universal high-frequency
  behavior of periodically driven systems: from dynamical stabilization to
  floquet engineering},}\ }\href
  {https://www.tandfonline.com/doi/full/10.1080/00018732.2015.1055918}
  {\bibfield  {journal} {\bibinfo  {journal} {Adv. Phys.}\ }\textbf {\bibinfo
  {volume} {64}},\ \bibinfo {pages} {139} (\bibinfo {year} {2015})}\BibitemShut
  {NoStop}%
\bibitem [{\citenamefont {Turitsyn}\ \emph {et~al.}(2012)\citenamefont
  {Turitsyn}, \citenamefont {Bale},\ and\ \citenamefont {Fedoruk}}]{TBFlasers}%
  \BibitemOpen
  \bibfield  {author} {\bibinfo {author} {\bibfnamefont {S.~K.}\ \bibnamefont
  {Turitsyn}}, \bibinfo {author} {\bibfnamefont {B.~G.}\ \bibnamefont {Bale}},
  \ and\ \bibinfo {author} {\bibfnamefont {M.~P.}\ \bibnamefont {Fedoruk}},\
  }\bibfield  {title} {\enquote {\bibinfo {title} {Dispersion-managed solitons
  in fibre systems and lasers},}\ }\href@noop {} {\bibfield  {journal}
  {\bibinfo  {journal} {Phys. Rep.}\ }\textbf {\bibinfo {volume} {521}},\
  \bibinfo {pages} {135} (\bibinfo {year} {2012})}\BibitemShut {NoStop}%
\bibitem [{\citenamefont {Biondini}(2008)}]{Biondini}%
  \BibitemOpen
  \bibfield  {author} {\bibinfo {author} {\bibfnamefont {G.}~\bibnamefont
  {Biondini}},\ }\bibfield  {title} {\enquote {\bibinfo {title} {The
  dispersion-managed {G}inzburg-{L}andau equation and its application to
  femptosecond lasers},}\ }\href@noop {} {\bibfield  {journal} {\bibinfo
  {journal} {Nonlinearity}\ }\textbf {\bibinfo {volume} {21}},\ \bibinfo
  {pages} {2849} (\bibinfo {year} {2008})}\BibitemShut {NoStop}%
\bibitem [{\citenamefont {Bale}\ \emph {et~al.}(2009)\citenamefont {Bale},
  \citenamefont {Boscolo}, \citenamefont {Schwartz},\ and\ \citenamefont
  {Turitsyn}}]{BBSTlocalized}%
  \BibitemOpen
  \bibfield  {author} {\bibinfo {author} {\bibfnamefont {B.~G.}\ \bibnamefont
  {Bale}}, \bibinfo {author} {\bibfnamefont {S.}~\bibnamefont {Boscolo}},
  \bibinfo {author} {\bibfnamefont {O.~Y.}\ \bibnamefont {Schwartz}}, \ and\
  \bibinfo {author} {\bibfnamefont {S.~K.}\ \bibnamefont {Turitsyn}},\
  }\bibfield  {title} {\enquote {\bibinfo {title} {Localized waves in optical
  systems with periodic dispersion and nonlinearity management},}\ }\href@noop
  {} {\bibfield  {journal} {\bibinfo  {journal} {Advances in Nonlinear Optics}\
  }\textbf {\bibinfo {volume} {21}},\ \bibinfo {pages} {181467} (\bibinfo
  {year} {2009})}\BibitemShut {NoStop}%
\end{thebibliography}
\end{document}